\documentclass[preprint,showpacs,preprintnumbers,amsmath,amssymb]{revtex4}

\usepackage{graphicx}
\usepackage{dcolumn}
\usepackage{bm}
\usepackage{amsmath}
\usepackage{amssymb}
\usepackage{latexsym}
\usepackage{epsfig}
\usepackage{amsbsy}
\usepackage{array}
\usepackage{amssymb}
\usepackage{setspace}
\usepackage{bm}

\def\sint{\ifmmode{- \!\!\!\!\!\! \int}
    \else{\hbox{$- \!\!\!\! \int \ $}}\fi}

\begin{document}


\title{ Single-pixel Tracking and Imaging under Weak Illumination }

\author{Shuai Sun$^{1,2}$}
\author{Hong-Kang Hu$^{1,2}$}
\author{Yao-Kun Xu$^{3}$}
\author{Hui-Zu Lin$^{1,2}$}
\author{Er-Feng Zhang$^{4}$}
\author{Liang Jiang$^{1,2}$}
\author{Wei-Tao Liu$^{1,2,*}$}

\affiliation{1, Department of Physics, College of Liberal Arts and Science, National University of Defense Technology, Changsha, Hunan 410073, China}
\affiliation{2, Interdisciplinary Center of Quantum Information, National University of Defense Technology, Changsha, Hunan, 410073, China}
\affiliation{3, College of Information and Communication, National University of Defense Technology, Xi'an, Shaanxi, 710106, China}
\affiliation{4, SSF Information Engineering University, Zhengzhou, Henan, 450002, China}
\affiliation{Corresponding author: wtliu@nudt.edu.cn}
\date{\today}


\begin{abstract}
    Under weak illumination, tracking and imaging moving object turns out to be hard. By spatially collecting the signal, single pixel imaging schemes promise the capability of image reconstruction from low photon flux. However, due to the requirement on large number of samplings, how to clearly image moving objects is an essential problem for such schemes. Here we present a principle of single pixel tracking and imaging method. Velocity vector of the object is obtained from temporal correlation of the bucket signals in a typical computational ghost imaging system. Then the illumination beam is steered accordingly. Taking the velocity into account, both trajectory and clear image of the object are achieved during its evolution. Since tracking is achieved with bucket signals independently, this scheme is valid for capturing moving object as fast as its displacement within the interval of every sampling keeps larger than the resolution of the optical system. Experimentally, our method works well with the average number of detected photons down to 1.88 photons/speckle.
\end{abstract}

\maketitle

\section{Introduction}
    Simultaneous tracking and imaging is a challenge for optical method, and become essential for many fields, such as biomechanics, on-line inspection and wearable AR/VR devices. Conventionally, moving object can be captured by sequential images produced by pixelated sensors at every transient moment\cite{Fuller2009, EI2009, Li2014, Naka2014, Gao2014}. Sensors of fast response and high sensitivity are necessary. An additional tracking and aiming system can help to relax such requirements, with additional complexity and the resolution limited by the tracking precision. In some scenarios, the photon flux arriving at the sensors within that transient moment is low due to scattering, absorption, low reflectivity of the object, or limited illumination power, etc. Then the conventional method will fail since the weak signal can not provide high enough signal to noise ratio for imaging or even tracking.

Ghost imaging (GI), as well as single pixel imaging (SPI), are capable to reconstruct image of the object with small photon flux \cite{Aspden2013, Morris2015, Aspden2015, Liu2018} by collecting the weak light into a single pixel detector, which usually possesses high sensitivity and fast response than array sensors. Besides, with the ability of interrogating at wavelengths outside the reach of array sensors, GI and SPI can achieve images at a wide range spectrum\cite{Yu2016, Zhang2016, Radwell2014, Olivieri2020, Stantchev2016}. However, the performance of GI and SPI for imaging moving objects is currently limited, since a large number of samplings are expected to be finished within the time interval that the objects can be treated as 'frozen'. Toward this, many approaches have been put forward to increase the sampling rate\cite{Kohno2019, Zhao2019, Zhou2020,Nitta2019,Xu2018} or reduce the required number of samplings \cite{Katz2009, Maga2013, Abmann2013, Phillips2017, He2018, Wang2019,Sun2019}. 
Benefits from the fact that tracking usually requires less information than spatially imaging, researchers also propose to acquire the trajectory of the moving object in lower dimension\cite{Shi2019} or in Fourier spectrum domain\cite{zzb2019}, with no image obtained. Considering the nature of GI, information are retrieved by correlation between fluctuations of the bucket signals and the illumination patterns. Such fluctuations can be introduced by the refreshing of the illumination patterns \cite{Gatti2004, Valencia2005, Chan2009, Shapiro2012, YK2018, Sli2018, Alba2019, WWL2019}, and can also be caused by motion of the objects\cite{Akhlaghi2017, Jiang2020}. With static or shifting spatial-resolved pattern, tracking was achieved from fluctuations produced by motion of the object \cite{Akhlaghi2017}. In GI and SPI, if the velocity or the trajectory is obtained, image of the object can be reconstructed gradually \cite{Jiang2020, SunS2019}. While, tracking and imaging under small photon flux is still an open problem.
        
In this paper, we demonstrate a single pixel tracking and imaging (SPTI) scheme within a typical computational GI (CGI) system. Velocity of the object is measured via temporal correlation of the bucket signal and the illumination beam is steered by a spatial light modulator(SLM) accordingly. Both image and trajectory of the object are recovered. Since tracking is achieved in an image-free way, our method is valid even if the displacement of the object within every sampling duration keeps larger than the spatial resolution of the system, as is far beyond the reach of existed SPI schemes. Collecting the photons with a photomultiplier, this method works well under low photon flux, with the average number of detected photons being as low as 1.88 photons/speckle. 

\section{Theoretical analysis}

For GI or SPI, to avoid the resolution degeneration, which is also referred to as motion blur, the data acquisition should be finished within the duration that the object can be treated as 'frozen'. To reconstruct a moving object from $N$ samplings, the transverse angular speed of the object $\omega(t)$ with respect to the imaging system is limited by  
\begin{equation}
    \int_0^{N/f}\omega(t)dt\leq \theta,
\end{equation}   
with $\theta$ and $f$ being the angular resolution and the sampling frequency of the system, respectively. That is, the transverse displacement during the $N$-sampling process should not be larger than the resolution of the system. Therefore the displacement during every individual sampling should be much smaller than the resolution, which strictly limits the speed of the object that can be captured. Besides, it also provides the requirement on the sampling frequency for GI or SPI schemes, which can be beyond the access of current technology if the moving object is too fast. We try to relax such limitations and advance SPI to SPTI, by tracking the object via temporal fluctuations of the bucket signals. 

In GI system, image of the object is reconstructed via fluctuation correlation as
      \begin{equation}
    G(\vec r_r)=\langle I(\vec r_r)B\rangle-\langle I(\vec r_r)\rangle \langle B\rangle,
      \end{equation}
   in which $\langle \cdot \rangle$ represents ensemble average over time or samplings. $I(\vec r_r)$ is the recorded pattern and $B$ is the bucket signals
     \begin{equation}
     B=\int T(\vec r_o) I(\vec r_o) d \vec r_o,
      \end{equation}
where $T(\vec r_o)$ is the reflectivity of the object and $I(\vec r_o)$ is the intensity distribution of the illumination pattern. In particular, under shifting illumination pattern, intensity of the bucket signal will fluctuate with the relative motion between the object and the pattern. Consequently, velocity of the object can be calculated by analyzing those bucket signals. Suppose the relative velocity is a constant within the interval that the illumination pattern keeps unchanged, the bucket result can be expressed as,
 \begin{equation}
    \begin{aligned}
        B(t)&=\int T(\vec r_o-\vec v_ot) I(\vec r_o-\vec v_i t) d \vec r_o \\
        & =\int T(\vec r_o) I(\vec r_o-\Delta \vec vt) d \vec r_o \\
        & = \int T(\vec r_o) I(\vec r_o-\Delta \vec r) d \vec r_o \\
    \end{aligned}
\end{equation}
  in which $\vec v_o$ is the transverse velocity of the object and the $\vec v_i$ is the shifting velocity of the illumination pattern. $\Delta \vec v=\vec v_i-\vec v_o$. $\Delta \vec r$ is the corresponding relative displacement. Considering the temporal correlation as
   \begin{equation}
  C(\tau)= \langle B(t)B(t+\tau)\rangle_{t_m},
   \end{equation}
   where $t_m$ is defined as velocity measurement interval, within which the pattern keeps unchanged and $N'$ samplings are performed. Substitute Eq. (4) into Eq. (5), we can obtain 
    \begin{equation}
   C(\tau)= \langle C_T(\Delta \vec r(t))C_I(\Delta \vec r(t))\rangle_{t_m},
   \end{equation}
   with $C_T(\Delta \vec r(t))$ and $C_I(\Delta \vec r(t))$ representing the autocorrelation function of the illumination pattern and that of the reflectivity of the object, respectively. With the relative displacement increasing, the overlapped area of the pattern imprinted on the object is reduced, causing the degree of correlation decreases. Thus the full width at half maximum (FWHM) of $C(\tau)$ is inversely proportional to $\Delta\vec r$. If the velocity of the object can be treated as constant within the measurement time, which is easy to achieve in practice, the FWHM of $C(\tau)$ will be inversely proportional to the relative velocity between the object and the illumination pattern. Therefore the relative velocity between the object and illumination pattern can be expressed as
    \begin{equation}
      |\Delta \vec v|\propto\frac{1}{\Delta \tau},
     \end{equation}
    where $\Delta \tau=\arg \frac{1}{2}{C(0)}$ is the FWHM of $C(\tau)$. To solve $\vec v_o$, we set the illumination pattern shifting in two orthognal directions with speed of $v_{x1}$, $v_{x2}$, $v_{y1}$, $v_{y2}$, seperately. Decomposing the velocity of the object as $\vec v_o=\vec v_{x}+\vec v_{y}$, four equations can be established and $v_x$ and $v_y$ can be solved, which means the velocity vector of the moving object is obtained.

According to the velocity of the moving object, the illumination beam can be steered to illuminate the object persistently. With enough number of samplings, the image of the object at instant $t$ can be reconstructed as,
      \begin{equation}
    G(\vec r(t))=\langle \mathcal{S}(\Delta \vec v(t) )I(\vec r(t))B(t)\rangle-\langle \mathcal{S}(\Delta \vec v(t))I(\vec r(t))\rangle \langle B\rangle,
      \end{equation}
where $ \mathcal{S}(\Delta \vec v(t) )$ represents the shifting operation on the reference patterns according to the relative velocity at every moment. To reduce the noise caused by imperfect illumination and the limited number of samplings, an enhanced correlation algorithm can also be used as\cite{SSun2019},
      \begin{equation}
    SG(\vec r(t))=\frac{\langle  \mathcal{S}(\Delta \vec v(t) )I(\vec r(t))B(t)\rangle\langle\int_\Omega  \mathcal{S}(\Delta \vec v(t) )I(\vec r(t)) d\vec r(t)\rangle}{\langle B\rangle \langle \mathcal{S}(\Delta \vec v(t) )I(\vec r(t)) \int_\Omega  \mathcal{S}(\Delta \vec v(t) )I(\vec r(t)) d\vec r(t)\rangle},
      \end{equation}
with $\Omega$ representing the region of the object. Then the position of the object at instant $t$ can be acquired from the reconstructed image. Combining the velocity at every moment, the trajectory of the moving object can also be recovered. 


With our scheme, both tracking and imaging can be achieved with a single pixel detector thus SPI scheme is advanced to SPTI. To solve the velocity, it is assumed that the velocity of the object can be treated as constant within $t_m$, which actually means
\begin{equation}
    \int_0^{N'/f}\frac {\partial \omega(t)}{\partial t}dt\ll \omega(t),
     \end{equation}        
with $f$ being the sampling frequency. In practice, several hundreds samplings are enough for velocity calculation with considerable accuracy. For a GI or SPI system with sampling rate of $10^5$Hz\cite{Xu2018, Zhao2019} or higher\cite{Kohno2019,Nitta2019,Zhou2020}, the velocity measurement time $t_m$ is in the order of microsecond, within which the velocity can be taken as constant for most cases. Compared with Eq. (1), the speed of the object is no more limited. Or, the requirement on sampling rate and the relationship between the spatial resolution and the maximum displacement is greatly relaxed. Even if the displacement of the object is far larger than the resolution within the sampling duration, our method can still work. 

\section{Experiments and Results}

A digram of the experimental setup is shown in Fig. 1. A fiber laser of 1064nm is cut into pulses of 20ns by an intensity modulator({Photline NIR-MX-LN-10}), with the repetition frequency of the pulses being 10kHz. The beam is enlarged via a 4-f system to match the spatial light modulator (SLM,Hamamatsu X13138-07). The speckle patterns are generated by controlled phase modulation on the SLM, which also performs beam steering. The object, a hex wrench of model $M2$ with the end face being illuminated, is driven by 2-dimensional motorized stages (Zolix PSA 100-11-X). The transverse coherence length of the illumination pattern on the object plane, which is 1m away from $L_3$, is about 88 mm thus the angle resolution of the system is 0.088mrad. The bucket signals from the object is collected by $L_4$ and detected by a photomultiplier tube(PMT). The phase masks to produce shifting illumination pattern is numerically obtained with the GS algorithm\cite{GS1972}. In our experiments, the sampling rate is set as $1$Hz, limited by the refresh frequency of the SLM. Those pulses modulated by the same phase screen will produces repetitive signals of the bucket detector, all of which is averaged as the final detection results.


\begin{figure}[h]
  \centering
  \includegraphics[width=\columnwidth]{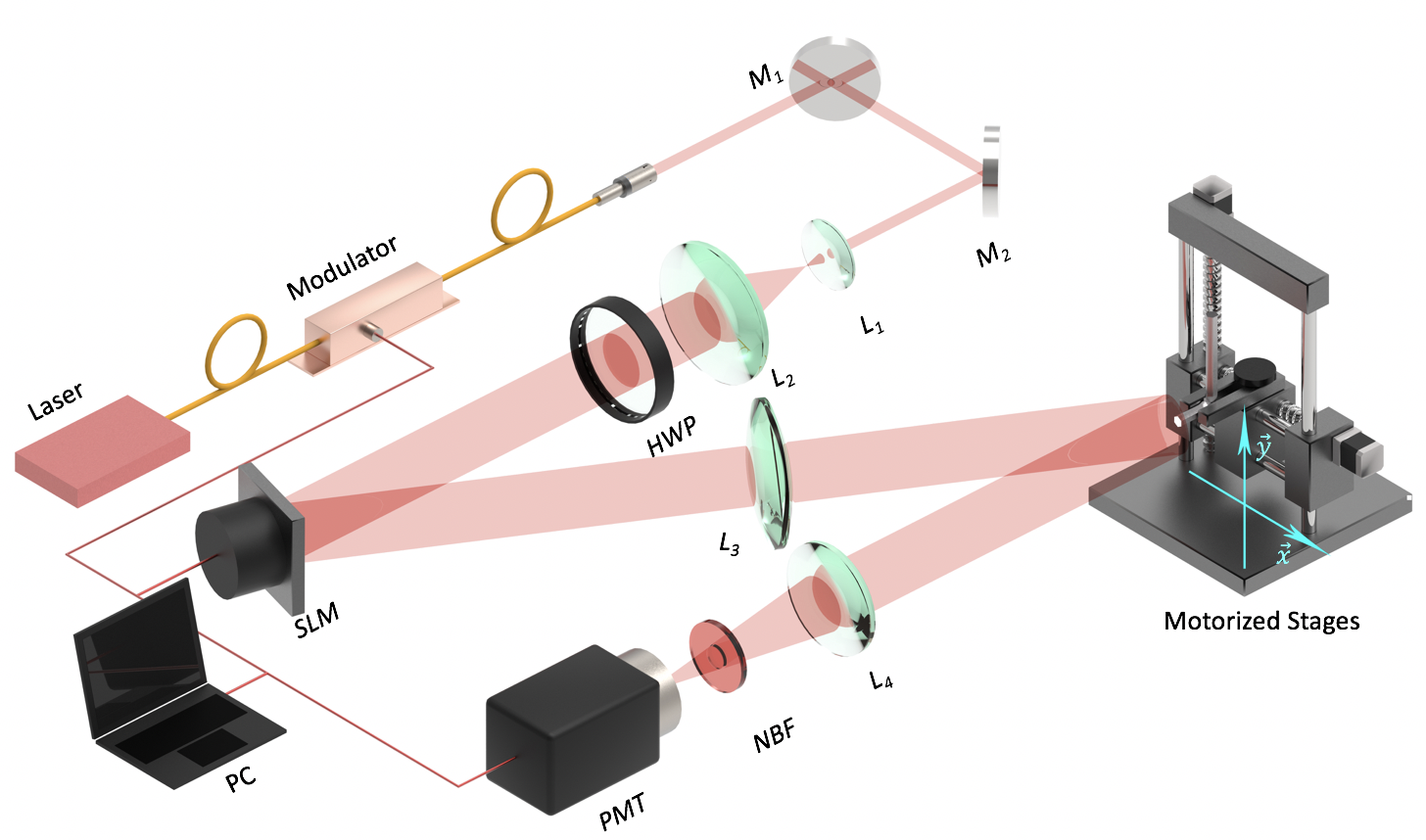}
  \caption{Experimental setup. Both $M_1$ and $M_2$ are mirrors. $L_1$ and $L_2$ consists a 4-f system, in which $f_1=5cm$ and $f_2=50cm$. HWP is a half-wavelength plate. NBF is narrow band filter with center wavelength of 1064nm and FWHM of 10nm. $f_3=100cm$ and $f_4=10cm$.}
  \label{setup} 
\end{figure} 

\subsection{Tracking via temporal correlation of bucket results}

Estimation of velocity and tracking is firstly tested for 1-dimension moving object, which is moving at a constant velocity $+0.05mrad/s$ in $\vec x$. Taking the anglular resolution and the duration of a single sampling as unit length and unit time, the speed of the object can also be expressed as 0.57 speckle/sampling. Under this speed, it is extremely hard for existed SPI/GI to capture the object since only less than two samplings can be performed during the interval that the object can be treated as 'frozen'. To estimate the velocity, two sequences of shifting pattern at different speed are illuminated on the object. The signals are collected by PMT, as depicted in Fig 2 (a), with $S_1$ and $S_2$ obtained under the velocity of the shifting pattern being $+0.10mrad/s$ and $+0.15mrad/s$, respectively. The temporal correlation of $S_1$ and $S_2$ are ploted in Fig. 2 (b), which shows $C(\tau)$ of $S_2$ decorrelates about twice as fast as that of $S_1$ since the relative velocity for $S_2$ is twice as large as that of $S_1$. With 1000 samplings, the velocity is calculated as $+0.047mrad/s$ (0.53speckle/sampling) according to Eq. (8). Tracking of moving objects at different speeds is also tested with the same number of samplings. The results are shown in Fig. 2 (c), with the black stars showing the real velocity and the circles showing the calculated values. Errors of the calculated velocity are mainly caused by limited number of statistics, thus can be reduced by increasing the number of samplings. With the object moving at $+0.1mrad/s$, the difference between the calculated speed and the real value is shown in Fig 2 (d), for different numbers of samplings. 

\begin{figure}[h]
  \centering
  \includegraphics[width=\columnwidth]{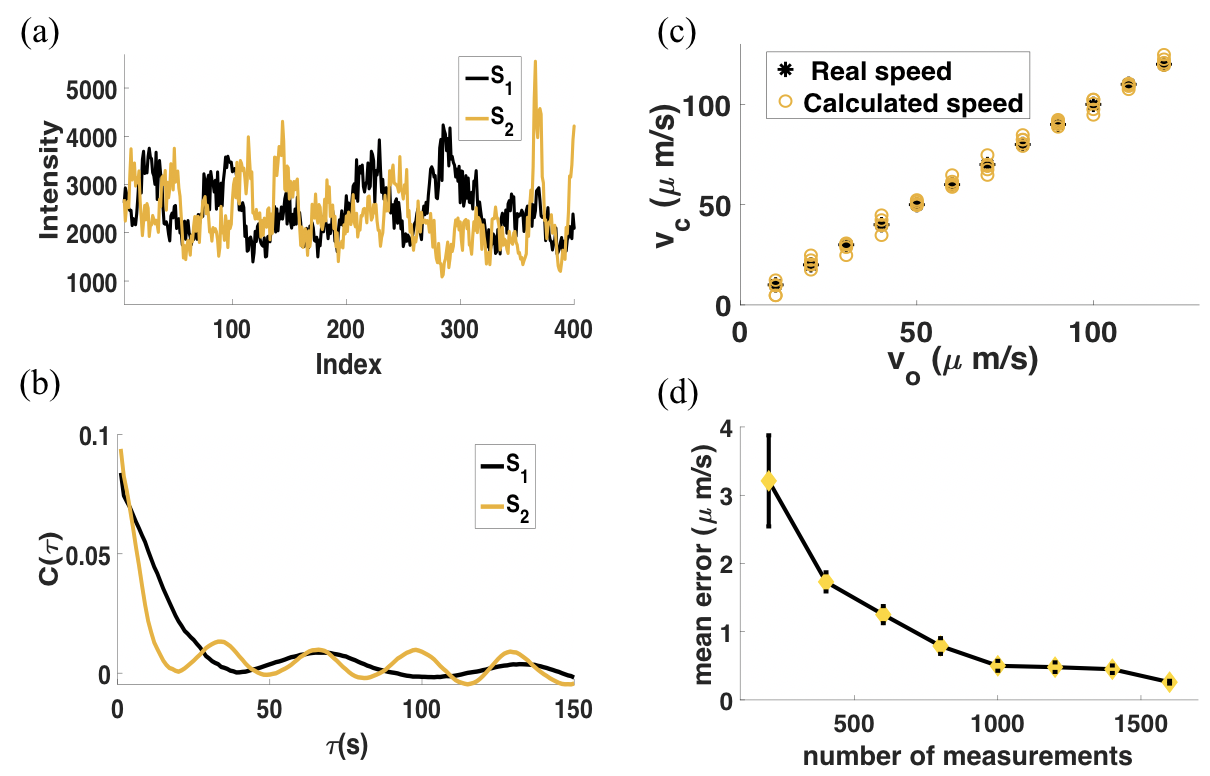}
  \caption{Verification of the validity of the velocity measurement from bucket signals. (a) shows the recorded bucket signals. (b) shows the temporal correlation of the bucket signals, the FWHM of which depend on the relative velocity between the object and the illumination pattern. (c) shows the real ($v_o$) and calculated speed ($v_c$) when the object is moving at different speed. (d) is the mean error between calculated and real speed.}
  \label{velocity} 
\end{figure} 

\subsection {Tracking and imaging of moving object}

Then the object is set moving faster than $0.1mrad/s$ (1.14speckle/sampling) at variable speeds in two dimension, with the trajectory depicted as the black line ($RT$) in Fig. 3 (a). The trajectory of the moving object consists of seven sub-paths, and the velocity of the object remains the same within each sub-path. The projections of the trajectory on $\vec x$ and $\vec y$ direction are shown in Fig. 3 (b). In each sub-path, four series of shifting illumination patterns are generated to measure the velocity of the object. Then the illumination beam is steered according to the calculated velocity, to ensure that the object is illuminated persistently. With 6400 samplings, both image and trajectory of the object are recovered, as shown in Fig. 3. 

It should be noted that, errors between the calculated and the real velocity of the object will cause accumulating errors in estimation of relative displacement, which will consequently affect the image reconstruction. Towards this issue, three steps are implemented to reduce the relative displacement error. First, the data sequence is divided into several segments, and a blurred image is reconstructed from each. Then the displacement of the object at different position is obtained via cross correlation between the blurred images, and a final image is recovered by superposition of the blurred images according to the displacement\cite{SunS2019}, with the results shown in the third column of Fig. 3 (d). Secondly, a better value of velocity is searched within the neighboring of the calculated value. Since smaller velocity error implies higher quality of image, quality of the final image is set as the target function for finding out velocity of smaller error \cite{LiE2014,Li2015,Jiao2019}. After that, error of the relative displacement got suppressed and an image with less noise was retrieved, as shown in the forth column of Fig. 3 (d). Thirdly, the enhanced correlation algorithm shown in Eq. (9) is used, with the result shown in the fifth column of Fig. 3 (d). As a comparison, image of static object is also interrogated with 6400 samplings and recovered via Eq. (9), shown in the last column. From the results, the quality of images of moving object, for example the resolution and the contrast, is comparable to the static one. The second column of Fig. 3(d) shows the results of traditional GI with the same number of samplings. The image of the object can not be extracted due to its movement.

With the recovered image, the location of the object at the final instant is acquired. Combining with the calculated velocity of the object at every moment, the trajectory is obtained and depicted as the orange line $ET_1$ in Fig. 3 (a). The error between the calculated and the real position in $\vec x$ and $\vec y$ direction is also shown. Although the position error is larger than the optical resolution at some moment, after the three steps optimization mentioned above, the image of the object with high quality can still be reconstructed by our method. 

To demonstrate that our method also works well in the case that the reflectivity of the object is not uniform, capturing of a screw, as shown in the leftmost column of Fig. 3 (e), is also performed. The trajectory is also depicted in the $RT$ in Fig 3 (a) and the recovered one is shown by $ET_2$. The projection of the calculated trajectory and that of the position error is shown in Fig. 3 (b) and Fig. 3 (c) with the same color, respectively. The second column of Fig. 3 (e) is the result from traditional GI, the third to fifth columns show the reconstructed image under the same three steps optimization, and the last column is the image of static object with the same number of samplings.

In this experimental setup, to match the condition shown in Eq. (10), the trajectory of the object is discretized and each step is associated with one velocity measurement, which costs about $10^3$ seconds due to the low refreshing rate of the SLM. Given higher sampling rate, the velocity measurement time can be reduced linearly and moving object with a smooth trajectory can be easily captured with our method. Besides, in this setup, using 5000 samplings to reconstruct an image for extraction of the object's position, those image-based tracking methods would be affected by motion blur if the speed of the object is larger than $\frac{0.088}{5000}mrad/s$ ($\frac{1}{5000}$ speckle/sampling), according to Eq. (1). By contrast, benefit from our image-free tracking method, moving object at a speed of larger than $0.1mrad/s$ (1.14speckle/sampling) is captured. 
\begin{figure}[h]
  \centering
  \includegraphics[width=\columnwidth]{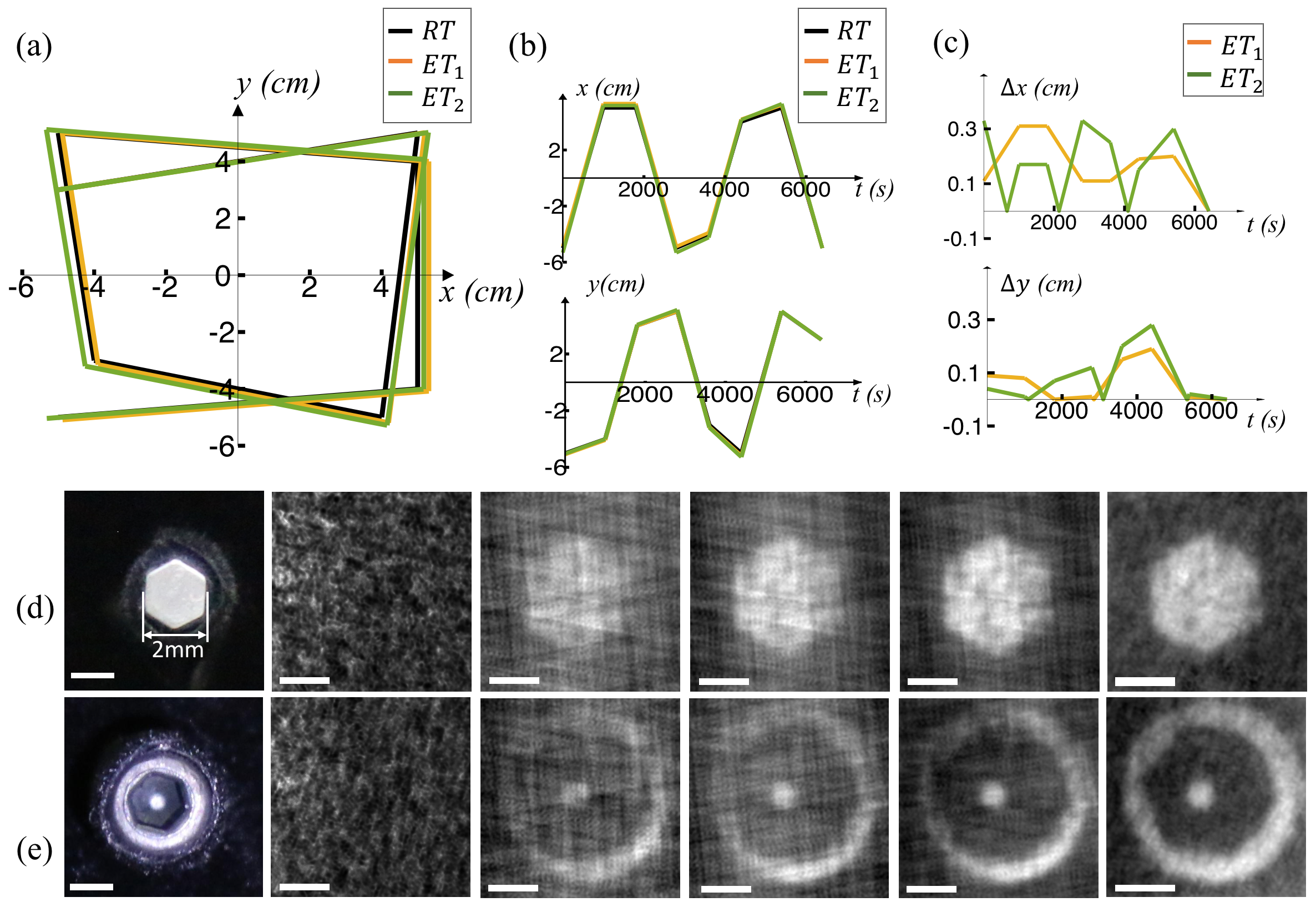}
  \caption {Capturing of object moving in two dimension. (a) depicts the real ($RT$) and the evaluated trajectory ($ET_1$ and $ET_2$) of the moving objects. (b) shows the projection of the real and the evaluated trajectory on the $\vec x$ and $\vec y$ direction. (c) shows the error between the evaluated and real trajectory in $\vec x$ and $\vec y$ direction. (d) and (e) shows the reconstructed images of the hex wrench and that of a screw, respectively, in which the leftmost column are pictures taken by camera, the second column is from traditional GI method, the third to fifth columns are captured images of moving object via our method and the last column is the image of static object with the same number of samplings as comparison. The white bar at the left bottom in the image is length scale.}
  
  \label{bright} 
\end{figure} 

\subsection{Performance at low photon flux}

To evaluate the performance of our method at weak photon flux, the repeat frequency of the laser pulse is decreased to 1Hz. The energy of a single pulse reflected from the SLM is about $1.69nJ$. Since the diffraction efficiency of the SLM is 0.55, the energy of the illumination pattern, which is produced by the 1st-order diffraction, is about $\frac{1.69nJ\times 0.55}{1+0.55}=0.60nJ$. The field of view is about $26mm\times 26mm$ and the area of the hex wrench is about $3.46mm^2$, from which the reflected photon flux is collected by $L_4$ located $1m$ away. The average current from the anode of PMT produced by a single pulse is ${\overline i}_s=17.13\mu A$ and the gain $\gamma$ is about $2\times10^3$, thus the average number of detected photons for a single pulse $\overline n$ can be estimated as 1070.59 ($\overline n=\frac{{\overline I}_s\tau_p}{\gamma e}$, in which $\tau_p$ is the duration of the pulse and $e$ is the charge of single electron). Since the average number of speckles imprinted on the object is about 569.55 (the ratio between the area of the object and the average area of the speckle), the average number of detected photon is about 1.88 photons/speckle, for each illumination pulse. 
Under this photon flux, if an array sensor of the same quantum efficiency is used to record the image of the object, 1.88 photons/pixel could be detected. Considering the shot noise and detection noise, the captured image will be very poor, which can not provide information for positioning and tracking of the object. Or, if trying to integrate over more than 2 pulses to increase the signal-to-noise ratio of the recorded image, motion blur will appear since the displacement of the moving object between two neighboring pulses is larger than the resolution. 

With our method, tracking and imaging is successfully performed. The object is moving along the trajectory depicted as black line in Fig 4 (a) with the projection on $\vec x$ and $\vec y$ direction shown in Fig 4 (b). The speed of the object is larger than $0.04 mrad/s$ (0.45speckle/sampling). With $1.4\times10^5$ samplings, the trajectory and image are recovered, with the results shown in Fig 4. Since the echo photons are collected with a single pixel detector and integration over time no more required, our method shows good performance under low photon flux detected. Moreover, since the velocity is recovered with correlation, uncorrelated detection noise is also suppressed. As a comparison, result from traditional GI is also shown in the leftmost column. In the rightmost column, image of static object with the same number of samplings and photon flux is shown.

\begin{figure}[h]
  \centering
  \includegraphics[width=\columnwidth]{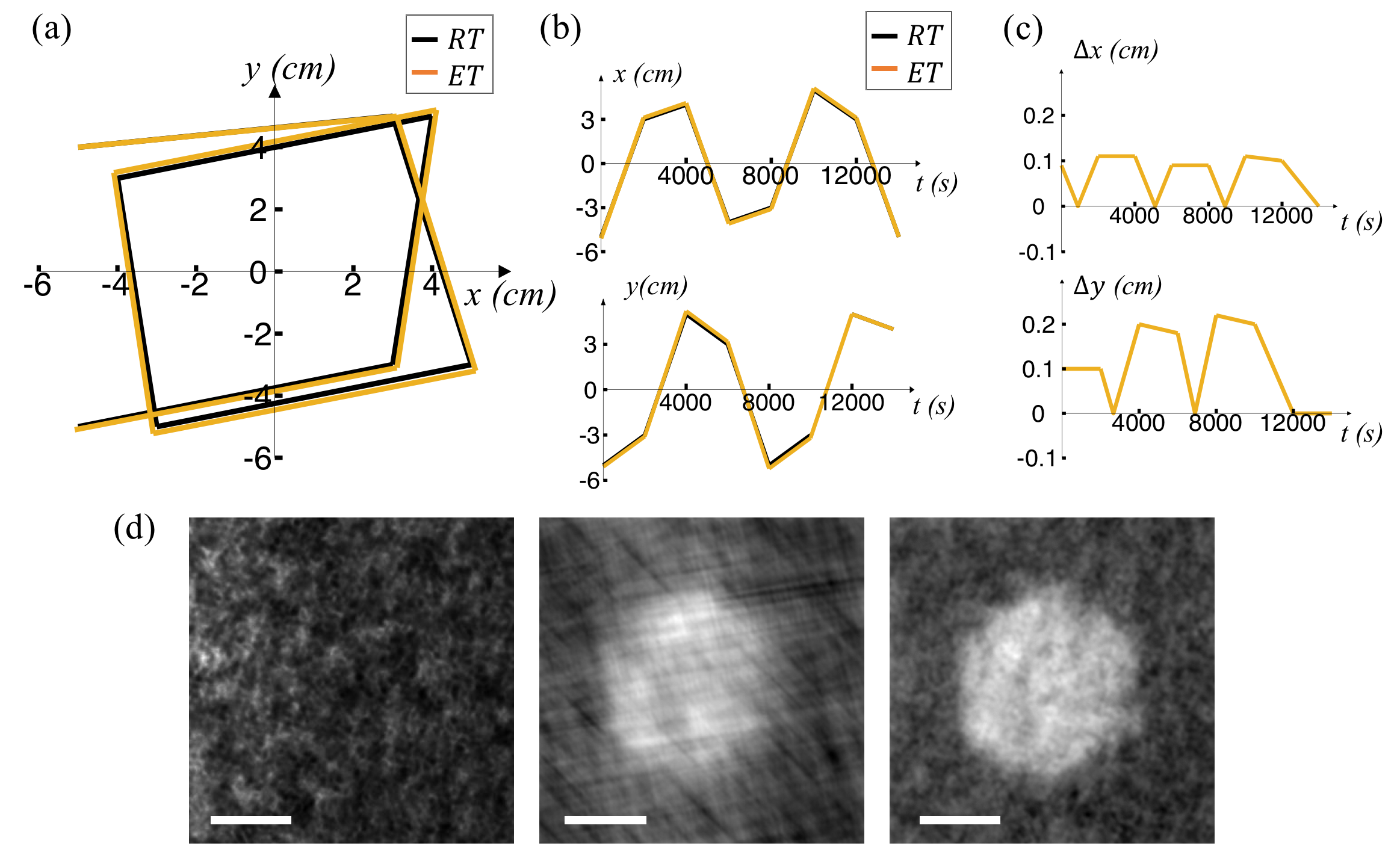}
  \caption{Capturing of moving object at low photon flux. (a) depicts the real (RT) and the evaluated trajectory (ET). (b) shows the projection of the real (RT) and evaluated trajectory (ET) on the $\vec x$ and $\vec y$ direction. (c) is the error between the evaluated and the real trajectory in $\vec x$ and $\vec y$ direction. The leftmost column is the results from traditional GI, the middle one shows the image of the moving hex wrench via our method and the last column is the image of static object with the same number of samplings. The white bar at the left bottom in the image is length scale.}
  \label{weak} 
\end{figure}  

\section{Further Discussion}
  In this scheme, error of the estimated velocity will be accumulated over time, which will deteriorate quality of the reconstructed image. Towards this issue, optimizations of three aspects were implemented in our experiments. In the case that the trajectory of the object is a more meandering curve, spline function interpolation\cite{Wernet1993,Pereira2006} can also be used to reduce the accumulation of the velocity error. On the other hand, the velocity of the object is obtained via correlation, in which the amplitude of fluctuation in the bucket detector fundamentally influences the error. Optimization in illumination design, considering the intensity distribution and transverse coherence length according to the scale of the object, can help to improve the calculation accuracy. Besides, increasing the number of sampling within the measurement time can always improving the signal noise ratio. In our method, the image of the object is reconstructed via intensity correlation shown by Eq. (9), which is a linear algorithm that can provide real-time imaging. Given longer time for data processing, the quality of image can be improved further using nonlinear algorithms\cite{Katz2009,Bian2018} or machine learning\cite{He2018,Wang2019}.
  
  In our experiments, limited by the refreshing rate of the SLM, the absolute speed of the moving object is not high. If the refresh rate of the illumination patterns is $10^5$Hz, moving object with angle speed of $10rad/s$ can be captured. For GI or SPI schemes to track and image moving objects, what matters essentially is the ratio between the displacement of the object and the resolution of the system within the duration of single sampling. Since the condition to capture moving object shown by Eq. (1) is replaced by Eq. (10), our method can work even if the displacement of the object within such duration is even larger than the resolution, as is far beyond the reach of existed GI and SPI schemes.  
  
  In practice, it is also common cases that the object moving in a complex or bright background, which is not considered in our experiment. In these situation, the bucket signals from the object can be obtained via background subtraction\cite{Sun2019}, then our method will be also effective to perform tracking and imaging.  
  
  Finally but not least, this method is promising in scenarios where only a small number of photons are achievable within every transient moment. Besides, considering the accessible spectrum range or the cost of the detector, our method can also perform tracking and imaging more economically in the scenarios that the operation wavelength is unsuitable for silicon-based sensor technology. These significant enhancements makes our method promising in autonomous vehicles, medical imaging and remote sensing applications.

\section*{Funding Information}
This work was supported by the National Natural Science Foundation of China under Grant Nos. 11774431, 61701511 and 61701537. W.T.Liu is supported by Science and Technology Project of Hunan Province (2017RS3043)

\end{document}